\documentclass[aps,prd,showpacs,superscriptaddress]{revtex4}
\usepackage{graphicx}
\usepackage{bm}
\begin{document}

\title{Correspondence between the adhesion model and
the velocity dispersion for the cosmological fluid}

\author{Takayuki Tatekawa}
\email{tatekawa@gravity.phys.waseda.ac.jp}
\affiliation{Department of Physics, Waseda University,
3-4-1 Okubo, Shinjuku-ku, Tokyo 169-8555, Japan}

\date{\today}

\begin{abstract}
Basing our discussion on the Lagrangian description of hydrodynamics,
we studied the evolution of density fluctuation for
nonlinear cosmological dynamics.
Adhesion approximation (AA) is known as a phenomenological model
that describes the nonlinear evolution of density fluctuation rather well
and that does not form a caustic. In addition to this model,
we have benefited from discussion of the relation between
artificial viscosity in AA and velocity dispersion.
Moreover, we found it useful to regard whether the velocity dispersion
is isotropic produces effective `pressure' or viscosity terms.
In this paper, we analyze plane- and spherical-symmetric cases
and compare AA with
Lagrangian models where pressure is
given by a polytropic equation of state.
From our analyses, the pressure model undergoes evolution
similar  to that of AA until reaching a quasi-nonlinear regime. 
Compared with the results of a numerical calculation,
the linear approximation of the pressure model seems rather good
until a quasi-nonlinear regime develops.
However, because of oscillation arising from the Jeans instability,
we could not produce a stable nonlinear structure.
\end{abstract}


\maketitle

\section{Introduction}\label{sec:intro}

The Lagrangian description for the cosmological fluid
can be usefully applied to
the structure formation scenario. This description provides a
relatively accurate model even in a quasi-linear regime.
Zel'dovich~\cite{zel} proposed
a linear Lagrangian approximation for dust fluid.
This approximation is called the Zel'dovich approximation
(ZA)~\cite{zel,Arnold82,Shandarin89,buchert89,coles,saco}. ZA
describes the evolution of density fluctuation better than
the Eulerian approximation~\cite{munshi,sahsha,yoshisato}.
Although ZA gives an accurate
description until a quasi-linear regime develops,
ZA cannot describe the model after the formation
of caustics. In ZA, even after the formation of caustics, the fluid elements
keep moving in the direction set up by the initial condition.
Therefore, the nonlinear structure that it is formed diffuses at once,
while N-body simulation shows the presence of
a dense structure
with a very wide range in mass at any given time~\cite{davis}.

In order to proceed with a hydrodynamical description
in which caustics do not form, the
`adhesion approximation'~\cite{gurbatov} (AA) was proposed
based on the model equation of nonlinear diffusion
(Burgers' equation). In AA, an artificial
viscosity term is added to ZA. Because of the viscosity term,
we can avoid caustics formation. From the standpoint of AA,
the problem of structure formation has been
discussed~\cite{Shandarin89,weinberg90,nusser,kofman,msw}.
The density divergence does not occur in AA, and the density distribution
close to the N-body simulation can be produced.
However, the origin of the viscosity has not yet been clarified.

Buchert and Dom\'{\i}nguez~\cite{budo} discussed the effect
of velocity dispersion using the collisionless Boltzmann equation~\cite{BT}.
They argued that models of a large-scale structure should
be constructed for a flow describing the average
motion of a multi-stream system.
Then they showed that when the velocity dispersion is regarded
as small and isotropic it produces effective `pressure' or
viscosity terms. Furthermore, they posited the
relation between mass density $\rho$ and pressure $P$, i.e.,
an `equation of state'.
Buchert et al.~\cite{bdp} showed how the viscosity term
or the effective pressure of a fluid is generated,
assuming that the peculiar acceleration is
parallel to the peculiar velocity.
Dom\'{\i}nguez~\cite{domi00,domi0106} clarified that a hydrodynamic
formulation is obtained via a spatial coarse-graining
in a many-body gravitating system, and that the viscosity term in
AA can be derived by the expansion
of coarse-grained equations.

With respect to the relation between the viscosity term
and effective pressure, and the extension of the Lagrangian description
to various matter,
the Lagrangian perturbation theory of pressure has been considered.
Actually, Adler and Buchert~\cite{adler} have formulated the
Lagrangian perturbation theory for a barotropic fluid.
Morita and Tatekawa~\cite{moritate} and Tatekawa et al.~\cite{tate02}
solved the Lagrangian perturbation equations for a polytropic fluid
up to the second order.
Hereafter, we call this model the `pressure model'.

In this paper, we analyze the evolution of the density fluctuation
in several Lagrangian models using simple models.
From these analyses, we examine the following questions.
(a) Can we explain the origin of the viscosity term in AA with
`pressure'?
(b) How long is the linear approximation of the pressure model
valid?
(c) Can we avoid the formation of a caustic with the pressure model?
To answer these questions, we analyze
time evolution for plane- and spherical-symmetric cases with first- (ZA),
second- (PZA), and third-order approximation (PPZA),
and for the exact solution for dust fluid, AA,
and the pressure model (linear approximation and full-order).
As shown by
previous papers~\cite{moritate, tate02, tate04}, the behavior
of the pressure model strongly depends on the polytropic exponent $\gamma$.
By the fine tuning of the parameter, the pressure model can reproduce
time evolution similar to that of AA until a quasi-nonlinear regime
develops. Furthermore, until the development of a quasi-nonlinear regime,
the linear approximation
of the pressure model seems rather good when compared with
a full-order numerical calculation.
However, the tendencies change greatly in a nonlinear regime.
Because of the Jeans instability, in the pressure model,
the density fluctuation oscillates. This oscillation of the fluctuation
appears in both the linear approximation and the full-order calculation.
Because the oscillation of the fluctuation does not occur in AA,
the pressure model cannot reproduce the behavior of AA completely.
Of course, as in the case of the dust fluid, the linear
approximation of the pressure model becomes worse in the nonlinear
regime.

The behavior of the fluctuation after the oscillation strongly depends
on the parameters.
In a previous paper~\cite{tate04}, although the case where $\gamma=5/3$
showed a rather good result when compared with N-body simulation,
caustics formation could not be avoided. Where
$\gamma=4/3$, the result seemed to resemble that of AA. However, in
a long-duration evolution of the pressure model,
even if a full-order equation was considered, caustics formed.
Where $\gamma=1$, the fluctuation disappeared.
Although behavior similar
to AA can be discussed with the pressure model
until a quasi-nonlinear regime develops, more consideration
is necessary to ascertain the existence of a stable nonlinear structure.

From our analyses, we conclude that we cannot sufficiently
explain the origin of the viscosity term in AA with the pressure model.
This conclusion does show, however, that we should apply the pressure model
in other situations. Recently, various dark matter models have been
proposed~\cite{DarkMatter}.
Some of them affect not only the gravity but also a special interaction.
We also show that the linear approximation of the pressure model
seems rather good until a quasi-linear regime develops.
If the interactions of the dark matter are affected by effective pressure,
the linear approximation can be applied for the analysis
of the quasi-nonlinear evolution of the density fluctuation.

This paper is organized as follows.
In Sec.~\ref{sec:Lagrangian}, we present Lagrangian perturbative solutions
in the Einstein-de Sitter (E-dS) universe.
In Sec.~\ref{subsec:dust}, we show perturbative solutions for dust fluid
up to a third-order approximation. Here we consider only the longitudinal mode.
In Sec.~\ref{subsec:adhesion}, we mention the problem of ZA and show
the solution of AA. In Sec.~\ref{subsec:pressure},
we explain the pressure model.

In Sec.~\ref{sec:compari}, we compare the evolution of the density
fluctuation between the Lagrangian approximations.
In Sec.~\ref{subsec:plane}, we analyze the plane-symmetric case.
Here ZA gives the exact solution for dust fluid.
In order to show the special tendency of this solution,
we analyze the spherical-symmetric case in
Sec.~\ref{subsec:spherical}.
In Sec.~\ref{sec:discuss}, we discuss our results
and state our conclusions.


\section{The Lagrangian description for the cosmological fluid}\label{sec:Lagrangian}

In this section, we present perturbative solutions in the Lagrangian
description.
In Lagrangian hydrodynamics,
the comoving coordinates $\bm{x}$ of the fluid elements are
represented in terms of Lagrangian coordinates $\bm{q}$ as
\begin{equation} \label{x=q+s}
\bm{x} = \bm{q} + \bm{s} (\bm{q},t) \,,
\end{equation}
where $\bm{s}$ denotes the Lagrangian displacement vector
due to the presence of inhomogeneities.
From the Jacobian of the coordinate transformation from
$\bm{x}$ to $\bm{q}$, $J \equiv \det (\partial x_i / \partial q_j)
= \det (\delta_{ij} + \partial s_i / \partial q_j)$,
the mass density is described exactly as
\begin{equation}\label{exactrho}
\rho = \rho_{\rm b} J^{-1} \,,
\end{equation}
where $\rho_{\rm b}$ means background average density.

We decompose $\bm{s}$ into the longitudinal
and the transverse modes as
$\bm{s} = \nabla_{\bm{q}} S
+ \bm{S}^{\rm T}$ with
$\nabla_{\bm{q}} \cdot \bm{S}^{\rm T}=0$. In this paper, we show
an explicit form of perturbative solutions only in the
Einstein-de Sitter (E-dS) universe.
%

\subsection{The Lagrangian perturbation for dust fluid}\label{subsec:dust}

Zel'dovich derived a first-order solution of the longitudinal mode
for dust fluid~\cite{zel}. For the E-dS model, the
solutions are written as follows:
\begin{equation} \label{eqn:sol-ZA}
S^{(1)} (\bm{q}, t) = t^{2/3} S_+ (\bm{q}) + t^{-1} S_- (\bm{q}) \,.
\end{equation}
This first-order approximation is called the Zel'dovich
approximation (ZA). Especially when we consider
the plane-symmetric case, ZA gives exact solutions
\cite{Arnold82}.

ZA solutions are known as perturbative solutions, which describe
the structure
well in the quasi-nonlinear regime. To improve approximation,
higher-order perturbative solutions of Lagrangian displacement
were derived.
Irrotational second-order solutions (PZA) were derived by
Bouchet et al.~\cite{bouchet92} and Buchert and Ehlers~\cite{bueh93},
and third-order solutions (PPZA) were obtained by Buchert~\cite{buchert94},
Bouchet et al.~\cite{bouchet95}, and Catelan~\cite{catelan}.
The second-order and third-order solutions are written
as follows:
\begin{eqnarray}
S^{(2)}_{i, i} &=& \frac{3}{14} \left(S^{(1)}_{i,j} S^{(1)}_{j,i}
 - S^{(1)}_{i,i} S^{(1)}_{j,j} \right ) \,, \label{eqn:sol-PZA} \\
S^{(3)}_{i, i} &=& \frac{5}{9} \left(S^{(1)}_{i,j} S^{(2)}_{j,i}
 - S^{(1)}_{i,i} S^{(2)}_{j,j} \right ) -\frac{1}{3} \mbox{det}
 \left (S^{(1)}_{i,j} \right ) \,, \label{eqn:sol-PPZA}
\end{eqnarray}
where the superscript $S^{(n)}$ means n-th order solutions.

Though third-order solutions have been obtained for
the transverse mode~\cite{buchert92,sasakasa}, because
we consider only longitudinal modes, we will pass over
those details here.

\subsection{Adhesion approximation}\label{subsec:adhesion}

Cosmological N-body simulations show that pancakes,
skeletons, and clumps remain during evolution. However, when we
continue applying the solutions of ZA, PZA, or PPZA after the appearance
of caustics, the nonlinear structure diffuses and breaks.

Adhesion approximation (AA)~\cite{gurbatov} was proposed from
a consideration based on Burgers' equation.
This model is derived by the addition of an artificial viscous term to ZA.
AA with small viscosity deals with the
skeleton of the structure, which at an arbitrary time is found directly
without a long numerical calculation.

We briefly describe the adhesion model. In ZA, the equation for
`peculiar velocity' in the E-dS model is written as follows:
\begin{eqnarray}
\frac{\partial \bm{u}}{\partial a} + (\bm{u} \cdot \nabla_x) \bm{u}
&=& 0 \,, \\
\bm{u} \equiv \frac{\partial \bm{x}}{\partial a}
 = \frac{\dot{\bm{x}}}{\dot{a}} \,,
\end{eqnarray}
where $a (\propto t^{2/3})$ means scale factor.
To go beyond ZA, we add the artificial viscosity term to the right side
of the equation.
\begin{equation} \label{eqn:adhesion}
\frac{\partial \bm{u}}{\partial a} + (\bm{u} \cdot \nabla_x) \bm{u}
= \nu \nabla_x^2 \bm{u} \,.
\end{equation}

We consider the case when the viscosity coefficient $\nu \rightarrow +0$
 ($\nu \ne 0$). In this case, the viscosity term especially affects the
high-density region. Within the limits of a small $\nu$, the analytic solution of
Eq.(\ref{eqn:adhesion}) is given by
\begin{equation}
\bm{u} (\bm{x}, t) = \sum_{\alpha} \left(\frac{\bm{x}-\bm{q}_{\alpha}}
{a} \right ) j_{\alpha} \exp \left( -\frac{I_{\alpha}}{2\nu} \right )
/ \sum_{\alpha} j_{\alpha} \exp \left( -\frac{I_{\alpha}}{2\nu} \right )
\,,
\end{equation}
where $\bm{q}_{\alpha}$ means the Lagrangian points that minimize the
action
\begin{eqnarray}
I_{\alpha} & \equiv & I(\bm{x}, a; \bm{q}_{\alpha})
 = S_0 (\bm{q}_{\alpha}) + \frac{(\bm{x}-\bm{q}_{\alpha})^2}{2a}
 = \mbox{min.} \,, \\
j_{\alpha} & \equiv & \left. \left[ \det \left (\delta_{ij}
 + \frac{\partial^2 S_0}{\partial q_i \partial q_j} \right )
 \right ]^{-1/2} \right |_{\bm{q}=\bm{q}_{\alpha}} \,, \\
S_0 &=& S(\bm{q}, t_0) \,,
\end{eqnarray}
considered as a function of $\bm{q}$ for fixed $\bm{x}$~\cite{kofman}.
In AA, because of the viscosity term, the caustic does not appear and
a stable nonlinear structure can exist.

\subsection{Pressure model}\label{subsec:pressure}
Although AA seems a good model for avoiding the formation of caustics,
the origin of the modification (or artificial viscosity) is not
clarified. Buchert and Dom\'{\i}nguez~\cite{budo} argued that the effect
of velocity dispersion becomes important beyond the caustics.
They showed that when the velocity dispersion is still
small and can be considered isotropic, it gives effective
`pressure' or viscosity terms.
Buchert et al.~\cite{bdp} showed how the viscosity term
is generated by the effective pressure of a fluid
under the assumption that the peculiar acceleration is
parallel to the peculiar velocity.

Adler and Buchert~\cite{adler} have
formulated the Lagrangian perturbation theory for a barotropic fluid.
Morita and Tatekawa~\cite{moritate} and Tatekawa et al.~\cite{tate02}
solved the Lagrangian perturbation equations for a polytropic fluid
in the Friedmann Universe.
Hereafter, we call this model the `pressure model'.

When we consider the polytropic equation of state $P=\kappa \rho^{\gamma}$,
the first-order solutions for the longitudinal mode are written as follows.
For $\gamma \ne 4/3$,
\begin{equation}\label{hatSbessel}
\widehat{S}(\bm{K},a) \propto a^{-1/4}
\, \mathcal{J}_{\pm 5/(8-6\gamma)}
\left( \sqrt{\frac{2C_2}{C_1}}
\frac{|\bm{K}|}{|4-3\gamma|}
\, a^{(4-3\gamma)/2} \right) \,,
\end{equation}
where $\mathcal{J}_{\nu}$ denotes the Bessel function of order $\nu$,
and for $\gamma=4/3$,
\begin{equation}\label{hatS43}
\widehat{S}(\bm{K},a) \propto
a^{-1/4 \pm \sqrt{25/16 - C_2 |\bm{K}|^2 / 2C_1}} \,,
\end{equation}
where $C_1 \equiv 4 \pi G \rho_{\rm b}(a_{\rm in})
\, a_{\rm in}^{\ 3} /3$
and $C_2 \equiv \kappa \gamma \rho_{\rm b}(a_{\rm in})^{\gamma-1}
\, a_{\rm in}^{\ 3(\gamma-1)}$. $\rho_b$ and $\bm{K}$ mean background
mass density and Lagrangian wavenumber, respectively. $a_{\rm in}$
means scale factor when an initial condition is given.
When we take the limit $\kappa \rightarrow 0$, these solutions
agree with Eq.~(\ref{eqn:sol-ZA}).

In this model, the behavior of the solutions strongly depends on
the relation between the scale of fluctuation and the Jeans scale.
Here we define the Jeans wavenumber as
\[
K_{\rm J} \equiv \left(
\frac{4\pi G\rho_{\rm b} a^2}
     {{\rm d} P / {\rm d} \rho (\rho_{\rm b})} \right)^{1/2} \,.
\]
The Jeans wavenumber, which gives a criterion for
whether a density perturbation with a wavenumber
will grow or decay with oscillation,
depends on time in general. If the polytropic index $\gamma$ is smaller
than $4/3$, all modes become decaying modes and the fluctuation
will disappear. On the other hand, if $\gamma > 4/3$, all density
perturbations will grow to collapse. In the case where $\gamma=4/3$,
the growing and decaying modes coexist at all times.

We rewrite the first-order solution Eq.~(\ref{hatSbessel}) with
the Jeans wavenumber:
\begin{equation}
\widehat{S}(\bm{K},a) \propto a^{-1/4}
\, \mathcal{J}_{\pm 5/(8-6\gamma)}
\left( \frac{\sqrt{6}}{|4-3\gamma|}
\frac{|\bm{K}|}{K_{\rm J}} \right) \,.
\end{equation}

In this paper, we analyze the first-order perturbation and
the full-order solution.
The evolution equation for the longitudinal mode is written as follows
~\cite{adler,moritate}:
\begin{equation} \label{eqn:P-full}
\nabla_x \cdot \left ( \nabla_q \ddot{\bm{s}} + 2 \frac{\dot{a}}{a}
 \nabla_q \dot{\bm{s}} - \frac{\kappa \gamma \rho_b^{\gamma-1}}{a^2}
 J^{-\gamma} \nabla_x J \right ) =-4 \pi G \rho_b (J^{-1} -1) \,.
\end{equation}
In general, it is very difficult to solve this equation for such
reasons as the coordinate transformation or non-locality.
Here, we imposed symmetry and avoided these difficulties.

\section{Comparison between Lagrangian models}\label{sec:compari}

\subsection{The plane-symmetric case}\label{subsec:plane}

First, we analyze the plane-symmetric case. In the plane-symmetric case,
ZA gives exact solutions for dust fluid. However, when we
keep using the solutions of ZA after the appearance
of caustics, the nonlinear structure diffuses and breaks.
We must connect the solutions with several procedures to continue
the calculation after the formation of caustics.

To simplify, we treat the single-wave case.
\begin{equation}
S_0(q)= \varepsilon \cos q \,.
\end{equation}
The initial peculiar velocity is made equal with that
given by the growing mode in ZA.
The evolution of this model for ZA, AA, and N-body simulation
(extrapolation of ZA) was analyzed by Nusser and Dekel~\cite{nusser}.
In this calculation, we set up the normalization of the scale factor
when first caustics appear with ZA by a=1. At a late time,
the caustics will diffuse in ZA.
AA remains a high-density filament
and caustics do not appear.

\begin{figure}
 \includegraphics{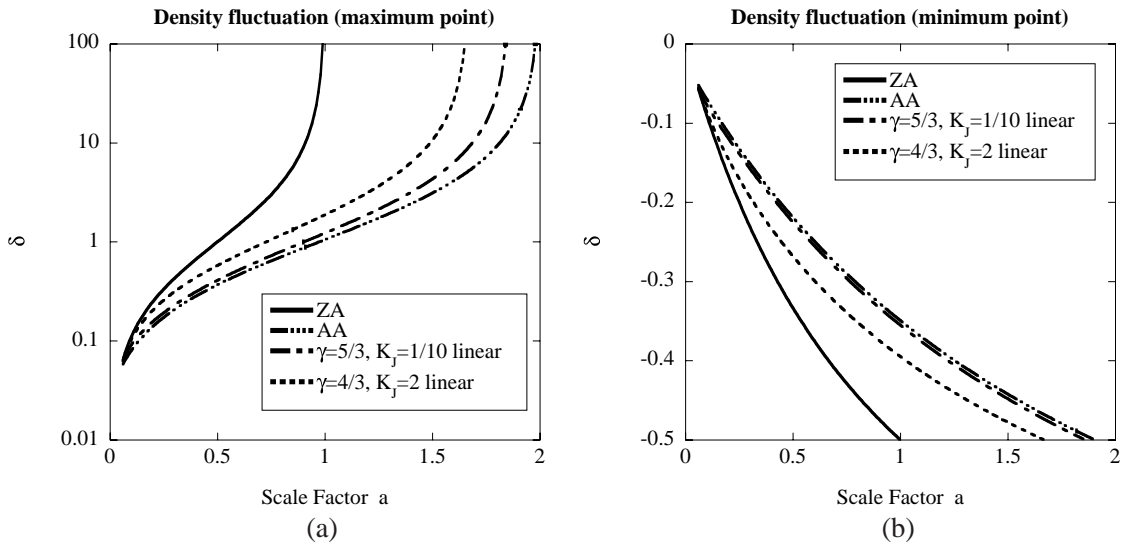}
 \caption{The evolution of density in linear approximations (ZA, AA, and
 the pressure model). The viscosity parameter in AA
 is given by $\nu=(\pi/512)^2$.
 (a) The density of the densest
 point. In AA, the density remains finite forever. On the other hand,
 in the pressure model, if $\gamma > 4/3$, the density becomes infinity.
 Therefore, we cannot avoid the formation of the caustic.
 (b) The density of the sparsest point. The growth of the void with
 the pressure model is fast in comparison with AA.}
\label{fig:1D-linear}
\end{figure}

\begin{figure}
 \includegraphics{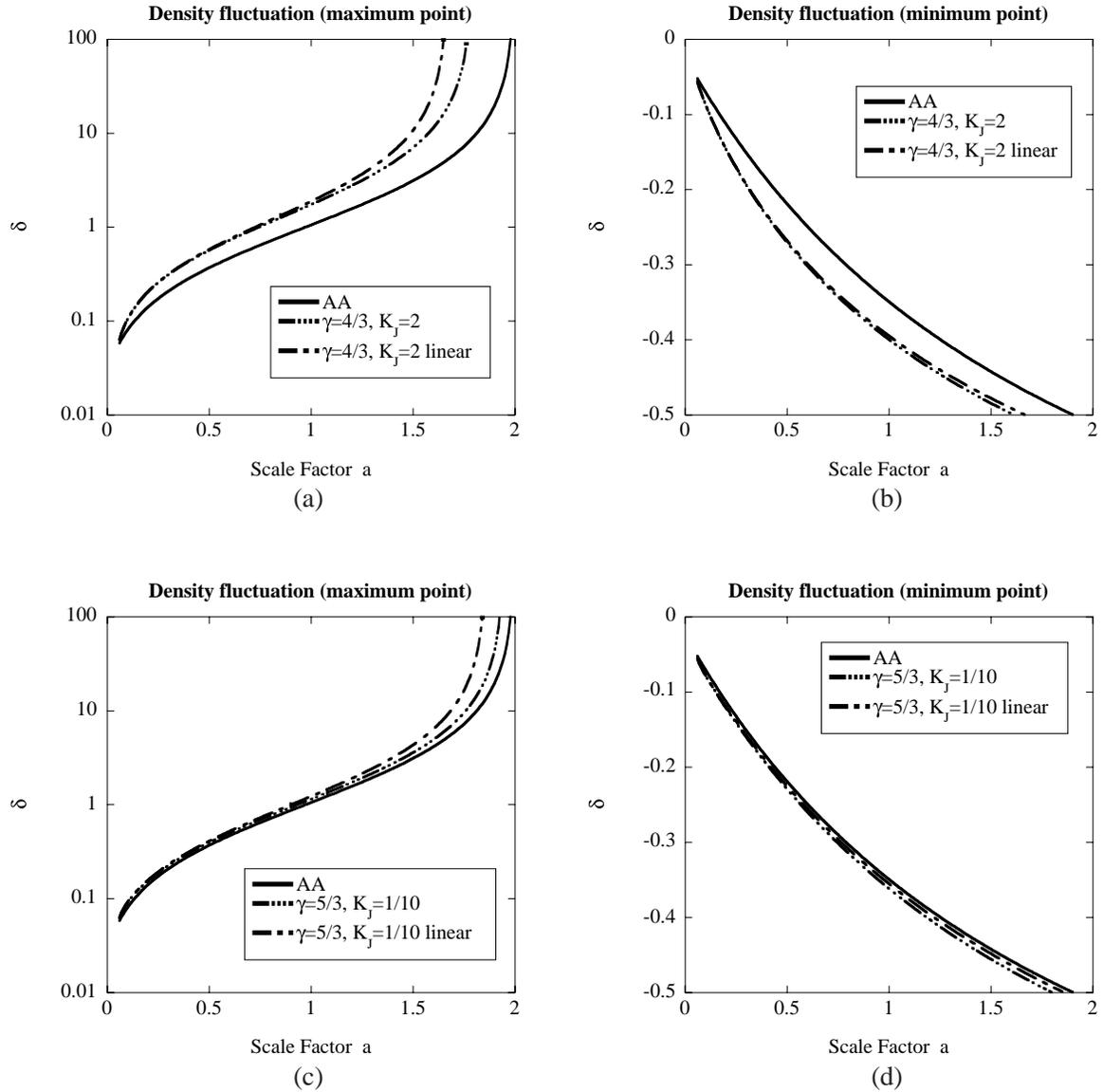}
 \caption{The evolution of density in AA and the pressure model
 (linear approximation and full-order). In the pressure model,
 the nonlinear effect suppresses the evolution of density fluctuation.
 Therefore,the linear approximation becomes worse in a strongly nonlinear
 regime. If $\gamma > 4/3$, although we calculate a full-order model,
 we cannot avoid the formation of the caustic.
 (a) The case where $\gamma=4/3$, the density of the densest point.
 (b) The case where $\gamma=4/3$, the density of the sparsest point.
 (c) The case where $\gamma=5/3$, the density of the densest point.
 (d) The case where $\gamma=5/3$, the density of the sparsest point.
 }
\label{fig:1D-Pressure}
\end{figure}

We analyze the time evolution of this model with the pressure models.
As mentioned in the previous section, the relation between the Jeans scale
and the scale of fluctuation is important for evolution in these models.
We consider only the case in which the scale of fluctuation is larger
than the initial Jeans scale. At first, we analyze the linear
perturbation of the pressure model.
In the linear perturbation, we consider the case where
$\gamma=4/3, 5/3$, because the fluctuation does not grow if
$\gamma < 4/3$.

Fig.~\ref{fig:1D-linear} shows
the evolution of density in ZA, AA, and the linear approximation
of the pressure model. As with the viscosity in AA,
the effect of the pressure delays the growth of the fluctuation.
In the case where $\gamma=5/3$, because the
perturbative solutions asymptotically become those of ZA,
the fluctuation grows rapidly at once.
In the case where $\gamma=4/3$, if we choose a reasonable value for $K_{\rm J}$,
though growth of the fluctuation can be slowed, caustics are
formed in the end (Fig.~\ref{fig:1D-linear} (a)).
In any case, though
it seems that the behavior of AA can be almost reproduced with
the pressure model by the fine-tuning of the parameter,
because the linear perturbative solutions keep
growing, the formation of the caustics cannot be prevented
where $\gamma > 4/3$. Therefore, the linear approximation
of the pressure model cannot reproduce the behavior of AA completely.
As for the region where the density fluctuation is negative
(Fig.~\ref{fig:1D-linear} (b)),
in comparison with ZA, AA and the pressure model suppress the
growth of the fluctuation.

Next we analyze the behavior of the solutions of the pressure
model without the approximation (Fig.~\ref{fig:1D-Pressure}).
Although the fluctuation keeps
growing when we use the linear approximation, we expect that the growth
of the fluctuation
may be restrained by the effect of nonlinearity. In fact,
in previous papers, we showed that the second-order perturbations
suppress the growth of the fluctuation~\cite{moritate,tate02}.

Here we analyze the case where $\gamma=4/3, 5/3$. In both cases,
the difference between the linear approximation deviates from the
full-order calculation greatly after $\delta > 1$.
In the case where $\gamma=4/3$ (Fig.~\ref{fig:1D-Pressure} (a), (b)),
though the behavior
of the solution strongly depends on the relation
between the scale of fluctuation and the Jeans scale,
we can delay the formation of caustics drastically.
However, when we analyze long-duration evolution, the density fluctuation
eventually diverges and the caustics form.
In the case where $\gamma=5/3$ (Fig.~\ref{fig:1D-Pressure} (c), (d)),
the growth of the fluctuation cannot be restrained considerably
either, as in the second-order
perturbation. Although good results were achieved in
the comparison with the N-body simulation~\cite{tate04}, it is
difficult to restrain the formation of the caustics in the case
where $\gamma=5/3$.

From these results, when $\gamma=4/3$, the growth of
the fluctuation can be gentle. However the caustics are finally formed,
the divergence of density cannot be avoided as with AA.
In other words, in the plane-symmetric case, we cannot represent
the behavior that resembles AA.

From Fig.~\ref{fig:1D-Pressure},
the linear approximation of the pressure model
gives a rather good result until a quasi-nonlinear regime develops.
In a strongly nonlinear regime,
the growth of the density fluctuation
in the linear approximation becomes slightly fast,
because of linearized pressure.

Because the results in this subsection may depend on symmetry,
we will analyze the spherical-symmetric case in the next subsection.

\subsection{The spherical-symmetric case}\label{subsec:spherical}

\begin{figure}
 \includegraphics{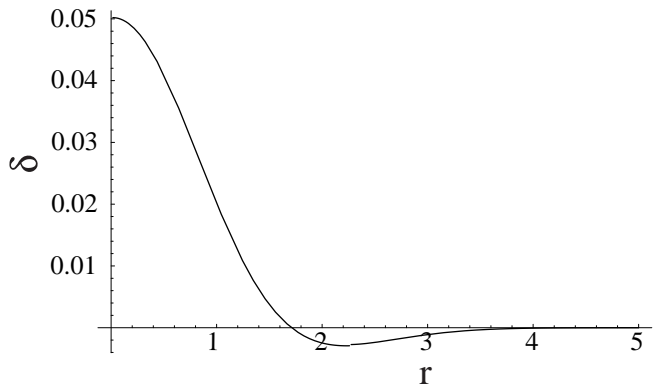}
 \caption{Mexican-hat type model. The average of density fluctuation
 over the whole space becomes zero.}
\label{fig:Mexican}
\end{figure}

For the spherical-symmetric case, dust collapse and void evolution
have been analyzed~\cite{munshi,sahsha,yoshisato}.
Here we consider the evolution with ZA, PZA, PPZA,
the exact solution for dust fluid, AA, and the
pressure models. To avoid a discontinuity of the pressure gradient,
we adopt the Mexican-hat type model (Fig.~\ref{fig:Mexican}):
\begin{equation}
\delta(r) = \varepsilon (3-r^2) e^{-r^2/2} \,.
\end{equation}
This model has several merits. For one the fluctuation is derived by
the two times differential calculus of Gaussian
\begin{equation}
- \nabla^2 \left ( \varepsilon e^{-r^2/2} \right )
 = - \frac{1}{r^2} \frac{\partial}{\partial r}
\left ( r^2 \frac{\partial}{\partial r}
 \left ( \varepsilon e^{-r^2/2} \right ) \right )
= \varepsilon (3-r^2) e^{-r^2/2} \,,
\end{equation}
and the average of density fluctuation over the whole space becomes zero:
\begin{equation}
\int_0^{\infty} 4\pi r^2 \delta(r) {\rm d} r = 0 \,.
\end{equation}

The initial peculiar velocity is made equal with that
of the growing mode in ZA.
For this model, from Eq.~(\ref{eqn:sol-ZA})-(\ref{eqn:sol-PPZA}),
the solutions of ZA, PZA, and PPZA are given as follows:
\begin{eqnarray}
S^{(1)} &=& - \varepsilon e^{-r^2/2} \,, \\
S^{(2)} &=& - \frac{3}{7} \varepsilon^2 e^{-r^2} \,, \\
S^{(3)} &=& - \frac{46}{189} \varepsilon^3 e^{-3 r^2/2} \,.
\end{eqnarray}

In our analysis, we set the value of $\varepsilon$ as follows:
\begin{equation}
\varepsilon = \pm \frac{1}{60} \,.
\end{equation}
Under this condition, the initial density fluctuation at $r=0$
becomes $\delta = \pm 0.05$. Then the scale factor is set as
$a=0.0167 (=1/60)$ at the initial condition.
In the case where $\varepsilon >0$, the caustics appear at $a=1$
in ZA.
The initial peculiar velocity is equal with that
given by the growing mode in ZA.

In past analyses~\cite{munshi,sahsha,yoshisato},
homogeneous spherical collapse and void evolution have been
analyzed. Here we consider spherical but inhomogeneous
density fluctuation. We investigate time evolution in the dust model
first because it may produce a result that differs from
that of past analyses.

\begin{figure}
\includegraphics{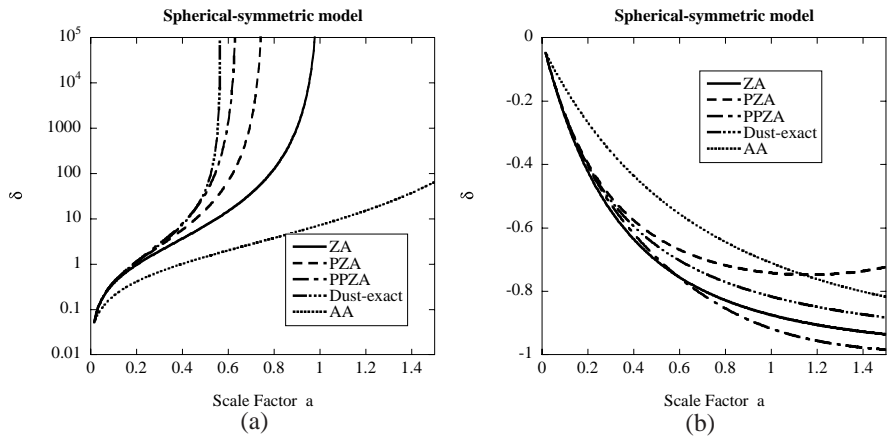}
 \caption{The evolution of the spherical-symmetric (the Mexican-hat
 type) density fluctuation at $r=0$ in dust models.
 (a) The evolution of a density fluctuation in ZA, PZA, PPZA,
 the exact model, and AA in the case where $\varepsilon >0$.
 The viscosity parameter in AA is set as
 $\nu = (1/512)^2$.
 The approximation is improved by higher order perturbation.
 In the exact model, the caustic appears
 at $a \simeq 0.55$. On the other hand, the density fluctuation
 evolves gently in AA.
 (b) The same as (a) but the case where $\varepsilon <0$.
 In PZA, the fluctuation becomes positive
 during evolution. Later ($a > 0.6$),
 PPZA deviates from an exact solution greatly more than ZA does.
 }
\label{fig:spherical-dust}
\end{figure}

Fig.~\ref{fig:spherical-dust} shows the time evolution of
the Mexican-hat type
density fluctuation in the dust model. For spherical collapse,
as well as in the past analyses,
when we considered higher order perturbation,
the occurrence time of the caustics
becomes fast~\cite{munshi,sahsha,yoshisato}.
The caustic appears with an exact solution at $a \simeq 0.55$.
On the other hand, the growth of the fluctuation becomes gentle,
and the caustic does not appear in AA.
For void evolution, the evolution
of the density fluctuation stops gradually with PZA,
and it starts to proceed in reverse.
When we consider long-time evolution, PPZA deviates
from an exact solution greatly more than ZA does.
These results correspond to past analyses considering
homogeneous spherical distribution.

\begin{figure}
\includegraphics{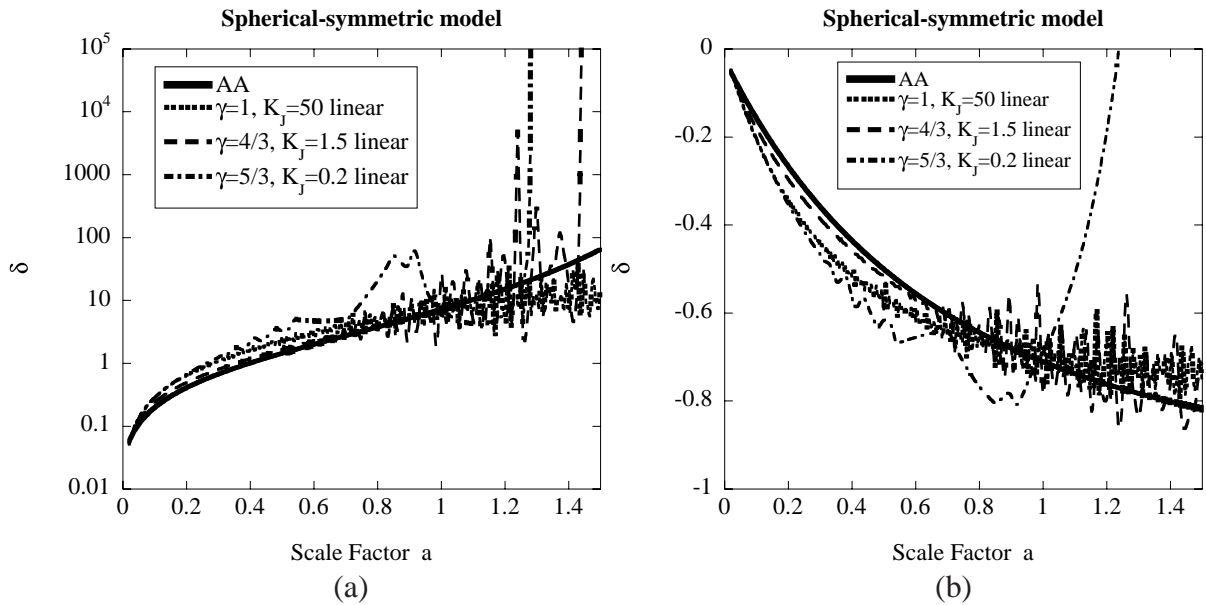}
 \caption{The evolution of the spherical-symmetric (Mexican-hat
 type) density fluctuation at $r=0$ in the pressure models
 with the linear approximation.
 (a) The evolution of a density fluctuation in the case where $\varepsilon >0$.
 Until a quasi-nonlinear regime develops,
 the pressure models show behavior resembling AA.
 However, the fluctuation oscillates in the nonlinear region.
 In the case where $\gamma=1$, the fluctuation oscillates little by little.
 Finally the fluctuation decays and disappears.
 In the case where $\gamma=4/3$, the fluctuation oscillates
 and the caustic appears at $a \simeq 1.44$.
 In the case where $\gamma=5/3$, the oscillation of
 the fluctuation in the intermediate state grows very large. Then
 the caustic appears at $a \simeq 1.28$.
 (b) The same as (a) but here $\varepsilon <0$.
 In the case where $\gamma=1$, finally the fluctuation decays and
 disappears.
 In the case where $\gamma=4/3$, although the fluctuation
 oscillates, the density asymptotically decreases.
 In the case where $\gamma=5/3$, the density fluctuation
 becomes positive during evolution because the oscillation of
 the fluctuation grows very large.}
\label{fig:spherical-linear}
\end{figure}

Next, we show how the Mexican-hat type fluctuation evolves
in the pressure model.
Fig.~\ref{fig:spherical-linear} shows the evolution
of the Mexican-hat type fluctuation in the pressure model
with linear approximation. Fig.~\ref{fig:spherical-linear} (a)
shows spherical collapse in the pressure models.
In the pressure model with linear approximation,
the evolution of the fluctuation shows strange
behavior. In the plane-symmetric case, the fluctuation includes only
the single wave mode. On the other hand, in the spherical-symmetric
case, the fluctuation includes various modes. Because of
the difference of the growth rate between the various modes,
the time evolution of the fluctuation does not become
monotonous.
At first, because we set the initial velocity in the
direction in which the fluctuation grows, the fluctuation grows
gently. Then, under the effect of pressure, the fluctuation begins
to oscillate. Finally, the fluctuation grows or decays.
The final state of the evolution strongly depends on the value of $\gamma$.

Here we adjust the value of $\kappa$, i.e., $K_{\rm J}$ in the pressure
models, to elicit behavior resembling a case of AA.
For the case where $\gamma=1$, when we consider the case of
a small Jeans scale, the fluctuation grows in the early stage.
After that, the fluctuation
oscillates little by little.
Finally, the fluctuation decays.
For the case where $\gamma=4/3$, as well as the
plane-symmetric case, the behavior depends on the
relation between the Jeans scale and the scale of the structure.
In linear approximation, when the scale of the structure is
larger than the Jeans scale, because the perturbative solution
produces the growing mode, the structure will collapse and form
caustics. In our analysis, when we choose $K_{\rm J}=1.5$, the
fluctuation oscillates and diverges.
The behavior of the fluctuation strongly depends on the value of $K_{\rm J}$.
Furthermore, relative to the growth rate of the fluctuation,
the time of the formation of caustics varies greatly over
a few differing values of $K_{\rm J}$.
For the case where $\gamma=5/3$, the oscillation of the fluctuation
in the intermediate state grows very large.
Then the caustic is formed.

Fig.~\ref{fig:spherical-linear} (b)
shows the void evolution in the pressure models.
As well as in the spherical collapse case,
the evolution of the fluctuation shows strange behavior.
When the fluctuation grows to $\delta \simeq -0.5$,
it begins to oscillate. Finally, the fluctuation grows or decays.
Especially, in the case where $\gamma=5/3$, the fluctuation shows
unrealistic evolution; the density fluctuation
becomes positive during evolution, because the oscillation of
the fluctuation grows very large.

When the fluctuation grew very large, we found that the behavior of AA could
not be reproduced any more in the pressure model with linear approximation.
In other words, it is very difficult to explain the origin of the viscosity
term in AA by the pressure model.
The oscillation of the fluctuation in the pressure model
with linear approximation is caused by the Jeans instability.
We will mention details about this oscillatory period and the amplitude
in Sec.~\ref{sec:discuss}.

\begin{figure}
\includegraphics{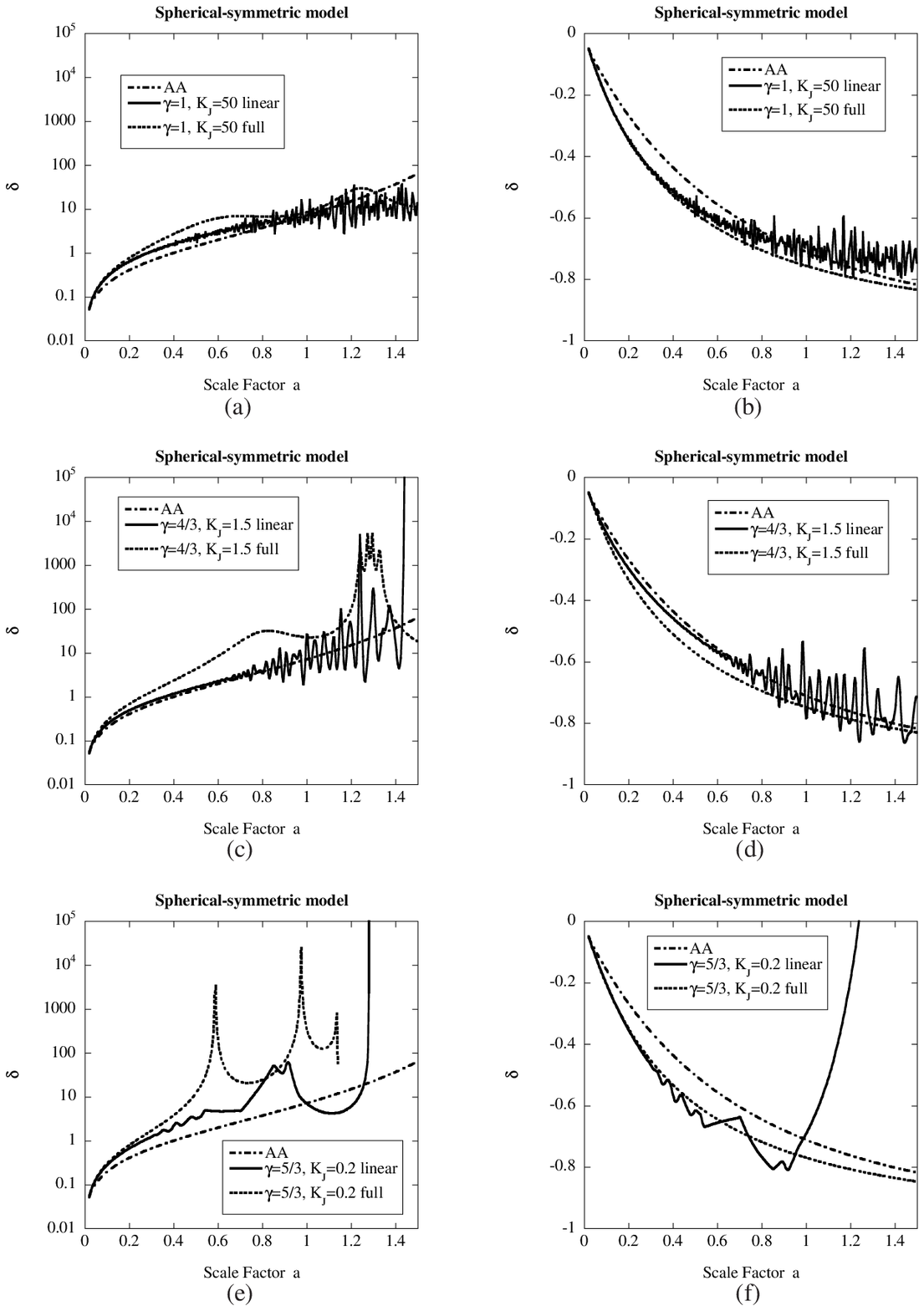}
 \caption{The evolution of a density fluctuation at $r=0$ in
 the spherical-symmetric case. These figures show the evolution
 in AA and the pressure models (linear approximation and
 the full-order calculation).
 These figures show the case where $\varepsilon >0$.
 (a) In the case where $\varepsilon >0$ and $\gamma=1$. In the pressure model,
 linear approximation seems valid until $\delta \simeq 1$.
 After that, in linear approximation, the fluctuation oscillates violently.
 In a strongly nonlinear region ($\delta >10$), even if we consider full-order
 calculation in the pressure model, the evolution of a fluctuation
 similar to AA cannot be reproduced.
 (b) The same as (a), but here $\varepsilon < 0$. As in the case where
 $\varepsilon >0$, when the fluctuation evolve fully ($\delta <-0.5$),
 the fluctuation begins to oscillate.
 Finally the fluctuation decays and approaches to $0$.
 (c) The same as (a), but here $\gamma=4/3$.
 In the pressure model with linear approximation,
 the fluctuation oscillates,
 and the caustic appears at $a \simeq 1.44$.
 When we consider a full-order equation,
 although we can delay the density divergence,
 we cannot avoid the formation of the caustic.
 (d) The same as (b), but here $\gamma=4/3$.
 In the linear approximation in the pressure model, although the fluctuation
 oscillates, the density asymptotically decreases.
 In the pressure model, when we consider a full-order calculation,
 we can realize the evolution of a fluctuation
 similar to that of AA.
 (e) The same as (a), but here $\gamma=5/3$.
 In the pressure model with the linear approximation, the oscillation of
 the fluctuation in the intermediate state grows very large. Then
 the caustic appears at $a \simeq 1.28$.
 When we consider a full-order equation, the density diverges a little
 to the outside at $a \simeq 1.14$, and the model fails.
 (f) The same as (b), but here $\gamma=5/3$.
 In the linear approximation in the pressure model, the density fluctuation
 becomes positive during evolution because the oscillation of
 the fluctuation grows very large.
 On the other hand, when we
 consider a full-order equation, it is different from linear
 approximation, the density fluctuation always remaining negative.
 }
\label{fig:3D-Full}
\end{figure}

When we solve the equation without approximation (Eq.~(\ref{eqn:P-full})),
how does the behavior of the density fluctuation change? These results
are shown in Fig.~\ref{fig:3D-Full}.
When we solve Eq.~(\ref{eqn:P-full}) for spherical collapse
($\varepsilon >0$) cases, such strange behavior
as violent oscillation is suppressed, and the evolution of
the fluctuation becomes smooth. However, these cases are different
from the plane-symmetric case; if pressure is ignored in
the spherical-symmetric case,
the gravity term contains non-linear terms.
Therefore, when we consider a full-order calculation,
the contribution of not only the pressure but also the gravity becomes
strong. Then if we choose a small value for $K_J$, the fluctuation
sometimes grows earlier than in the case of linear approximation
(Fig.~\ref{fig:3D-Full} (a), (c), and (e)).
In either case, the general tendency of the evolution of the fluctuation
does not differ very much.
According to Fig.~\ref{fig:3D-Full} (a), (c), and (e),
linear approximation of the pressure
model seems to good until the quasi-nonlinear regime develops.
The final state is unchanged, though a few differences are seen in
the growth of the fluctuation, oscillatory amplitude, and period.
In other words, even if the full-order calculation is considered,
it is very difficult to explain the origin of the viscosity
term in AA by the pressure model.

Next, we mention the evolution of a void (the case where $\varepsilon <0$).
When we consider a full-order calculation, unrealistic
behavior, such as linear approximation in the case where $\gamma=5/3$
(Fig.~\ref{fig:spherical-linear} (b)),
does not appear. Furthermore,
the oscillation of the fluctuation is suppressed, and the growth
of the fluctuation comes to look like that of AA.
In the evolution of the void, the linear approximation of the pressure
model seems good until $\delta \simeq -0.5$.
When we do not introduce linear approximation, the oscillation
of the density fluctuation is almost imperceptible
(Fig.~\ref{fig:3D-Full} (b), (d), and (f)).

Although we can realize a void evolution in AA with the pressure model,
we cannot reproduce the existence of a stable nonlinear structure.
In other words, it is very difficult to
find the origin of artificial viscosity in AA with the
isotropic velocity dispersion.

According to our calculation in linear approximation,
the amplitude and the period of
the oscillation of the fluctuation in the intermediate state
obviously depends on $\gamma$. Although the tendency of the
evolution of the fluctuation in the case of $\gamma=4/3$ looks
like AA, the snapshot of the density field will be different from
that in AA. We will mention the reason in discussion.

As for the validity of the linear approximation in the pressure model,
as well as the case of the plane-symmetric case,
the approximation is rather good until a quasi-nonlinear regime
develops.
However, attention is necessary for extrapolation to a nonlinear stage
with Lagrangian linear perturbation because it is different
from the plane-symmetric case, the oscillation of the density fluctuation
appearing in the spherical-symmetric case at the nonlinear regime.

\section{Discussion and Concluding Remarks}
\label{sec:discuss}

We analyzed the corresponding relation with the viscosity
term in AA and the velocity dispersion using plane- and
spherical-symmetric cases. Here we evaluated the effect
of isotropic velocity dispersion by linear approximation or
the full-order equation. As shown by
our previous papers~\cite{moritate, tate02, tate04},
we derived the basic equation in Lagrangian description.
We called this model the pressure model. The behavior
of the pressure model strongly depends on the equation of state.
Using AA, we can avoid the formation of the caustic, i.e. density
divergence. We studied carefully whether
a stable nonlinear structure could exist in the pressure model.
In our previous paper~\cite{tate04}, although the case where $\gamma=5/3$
showed a rather good result when compared with N-body simulation,
this case cannot avoid caustics formation. In the case where
$\gamma=1, 4/3, 5/3$, the result seems to resemble that in AA until
a quasi-nonlinear regime develops.
However, in long-duration evolution, even if we consider full-order effects,
the caustics will be formed. Though behavior similar to that of
AA can be seen with the pressure model, more consideration
is necessary for establishing the existence of stable nonlinear
structure.

Here we mention the reason why density fluctuation oscillated in case of
linear approximation with the pressure model. We also describe
the origin of the amplitude and the period of the oscillation.
The solution of the linear perturbation in pressure models are given
by Eq.~(\ref{hatSbessel}) and (\ref{hatS43}). In the case of
$\gamma=1, 5/3$,
i.e., $\nu= \pm 5/2$, the Bessel functions can be written with
trigonometric functions.
\begin{eqnarray}
J_{5/2} (z) &=& \sqrt{\frac{2}{\pi z}} \left \{ \left (
\frac{3}{z^2} - 1 \right ) \sin z - \frac{3}{z} \cos z \right \}
\,, \\
J_{-5/2} (z) &=& \sqrt{\frac{2}{\pi z}} \left \{ \frac{3}{z}
\sin z + \left ( \frac{3}{z^2} - 1 \right ) \cos z \right \} \,.
\end{eqnarray}
For the case where $\gamma=1$, the leading term of the solutions for
large $t$ becomes as follows:
\begin{equation}
D^+ \sim t^{-1/3} \sin (A |\bm{K}| t^{1/3}),
~~ D^- \sim t^{-1/3} \cos (A |\bm{K}| t^{1/3})
\,,
\end{equation}
where $A$ means constant.
On the other hand, for the case where $\gamma=1$, the leading term of
the solutions for large $t$ becomes as follows:
\begin{equation}
D^+ \sim t^{2/3} \sin (A |\bm{K}| t^{-1/3}) ,
~~ D^- \sim t^{2/3} \cos (A |\bm{K}| t^{-1/3})
\,.
\end{equation}
Therefore, in the case where $\gamma=1$, the amplitude of the fluctuation
decreases, and the period of the oscillation becomes relatively
short. On the other hand, in the case where $\gamma=5/3$, the amplitude
of the fluctuation grows like that of ZA, and the period of
the oscillation is prolonged.

For the case where $\gamma=4/3$, if the scale of the fluctuation is
smaller than the Jeans scale, the fluctuation oscillates. In this case
the linear perturbative solution is written as follows:
\begin{equation}
D^{\pm} \sim t^{-1/6 \pm A'i} \sim t^{-1/6} \cos (A' \log t),~
t^{-1/6} \sin (A' \log t) \,,
\end{equation}
where $A'$ means constant.
Therefore, the oscillation of the fluctuation is slower than
that in the case where $\gamma=1$. Then, the amplitude of the
fluctuation is smaller than that in the case where $\gamma=5/3$.
When we consider a full-order calculation, the oscillation of
the density fluctuation becomes gentle. The reason seems
to be mode-coupling in nonlinear evolution. However, this effect
cannot control the oscillation well; as we show in
Fig.~\ref{fig:3D-Full},
the oscillation remains.

If we adjust the parameters of the pressure model, will it be
able to obtain a result similar to the behavior of AA?
According to our work, it seems quite difficult to establish
the existence of a stable nonlinear structure. For example,
if we choose a small $\gamma$, although we can avoid the caustics
formation,
the fluctuation will decay and disappear. On the other
hand, if we choose a large $\gamma$, the fluctuation behaves
like that of ZA. Therefore, the fluctuation forms a caustic.
If the parameters
are chosen carefully, we may solve the problem of caustic formation
and fluctuation disappearance. However,
the oscillation of the fluctuation remains, even if we can realize
the tendency of the growth of the fluctuation.
If we hope to clarify the origin of artificial
viscosity in AA,
we need to consider other effects, for example,
spatial coarse-graining~\cite{domi00,domi0106},
anisotropic velocity dispersion~\cite{mtm}, and so on.
In future, we will analyze the nature of the model from which
the other effect was taken.

Next, we consider another question. When we analyze structure formation
in the fluid with pressure, can we learn whether the Lagrangian
linear perturbation is valid or not? From our analyses
in the plane- and spherical-symmetric cases,
until a quasi-nonlinear regime develops, the linear approximation
of the pressure model seems rather good from the comparison with
a full-order numerical calculation.
Therefore, for example, if the interaction in some kind of
dark matter can be described by the effective pressure, we can examine
the behavior of the density fluctuation in a quasi-nonlinear stage.
Furthermore, when we compare the observations and the structure
that is formed by using the pressure model, we can give
a limitation to the nature of the dark matter.

\begin{acknowledgments}
We are grateful to Kei-ichi Maeda for his continuous encouragement.
We would like to thank Thomas Buchert, Aya Sekido,
and Hajime Sotani for
useful discussion and comments regarding this work.
We would like to thank Peter Musolf for checking of
English writing of this paper.
Our numerical
computation was carried out by Yukawa Institute computer faculty.
This work was supported in part by a Waseda University Grant
for Special Research Projects (Individual Research
2003A-089).
\end{acknowledgments}

\end{document}